\documentclass[twocolumn,prd,secnumarabic,nobalancelastpage,amsmath,amssymb,
nofootinbib]{revtex4-2}

\usepackage{graphicx} 
\usepackage{xcolor}
\newcommand{\mL}{{\mathcal{L}}}
\newcommand{\mE}{{\mathcal{E}}}
\newcommand{\req}[1]{Eq.\,(\ref{#1})} 
\newcommand{\rs}[1]{section~\ref{#1}} 
\newcommand{\rf}[1]{Fig.\,\ref{#1}} 

\begin{document}

\preprint{arXiv:2108.12959v2 [hep-ph]}

\title{Particle production at a finite potential step: \\
Transition from Euler-Heisenberg to Klein paradox}

\author{Stefan Evans}
\email{evanss@email.arizona.edu}
\author{Johann Rafelski}
\affiliation{Department of Physics, The University of Arizona, Tucson, AZ 85721, USA}
\date{arXiv:2108.12959v2 [hep-ph], October 22, 2021}

\begin{abstract}
Spontaneous pair production for spin-$1/2$ and spin-$0$ particles is explored in a quantitative manner for a static $\tanh$-Sauter potential step (SS), evaluating the imaginary part of the effective action. We provide finite-valued per unit-surface results, including the exact sharp-edge Klein paradox (KP) limit, which is the upper bound to pair production. At the vacuum instability threshold the spin-$0$ particle production can surpass that for the spin-$1/2$ rate. Presenting the effect of two opposite sign Sauter potential steps creating a well we show that spin-$0$ pair production, contrary to the case of spin-$1/2$, requires a smoothly sloped wall. 
\end{abstract}

\maketitle

\section{Introduction}

We study exact rates for spontaneous particle production in strong electromagnetic (EM)-fields, signaling vacuum instability. Our work relies on analytical solutions of the relativistic wave equations~\cite{Sauter:1932gsa}: Sauter in 1932 employed a $\tanh$-like smoothed potential edge, the Sauter step (SS). Our results presented here will influence the development of experiments aiming to measure these processes in (non-relativistic~\cite{Rafelski:2016ixr,Popov:2020xmd}, and ultra-relativistic~\cite{Bertulani:2005ru,Tuchin:2013ie}) heavy ion collisions, as well as ultra-short pulsed lasers~\cite{Mourou:2006zz,DiPiazza:2011tq}, and help better understand related topics in astrophysics~\cite{Ruffini:2009hg,Churazov:2020}.

A better known Sauter solution obtained for constant electrical fields~\cite{Sauter:1931zz} laid the foundation for the understanding of pair production by Euler-Heisenberg-Schwinger (EHS)~\cite{Heisenberg:1935qt,Weisskopf:1996bu,Schwinger:1951nm,Dunne:2008kc,Dunne:2009gi}. This exact EHS result leads to a locally constant field approximation (LCF), {\it i.e.} frequent in literature use of EHS results for fields varying in space, disregarding field gradients. We will demonstrate quantitatively limitations of this approach. 

Sauter aimed in his SS work to improve the understanding of the \lq Klein paradox\rq\ (KP)~\cite{Klein:1929zz,Dombey:1999id}, reproduced by the SS in the sharp-edge limit with potential step width measured in terms of the particle Compton wavelength ($\lambdabar_C$) scale. The Klein paradox of 1929 was understood in the ensuing decades to be due to spontaneous nonperturbative pair production occurring when the SS potential height exceeds the pair production threshold $2m$, a vacuum instability in presence of strong fields; for the case of a finite domain in a space-time potential-well, this instability implies the spontaneous transition into a locally modified, charged vacuum state~\cite{Rafelski:1974rh, Rafelski:1976ts,Greiner:1985ce,Rafelski:2016ixr,Sveshnikov:2019aqg,Popov:2020xmd}. 

We are not aware that a connection between the two domains of strong field physics: The (constant field) EHS (called often \lq Schwinger\rq) pair production, and the `positron' production from local supercritical potential wells, has been clarified completely. By exploring in depth the SS-potential pair production, extending to the context of of SS-well, we reunite these two physics domains. We further demonstrate that the sharp KP limit requires a very large departure from the EHS particle production yields, and a change in thinking away from particle production per unit volume to particle production per unit (Sauter step) surface. Our work builds on the insights obtained between 1970 and the present day~\cite{Nikishov:1970br, Nikishov:1970bN,Nikishov:2003ig,Kim:2009pg, Chervyakov:2009bq, Chervyakov:2018nr, Gavrilov:2015yha}.

In 1970, Nikishov~\cite{Nikishov:1970br,Nikishov:1970bN} obtained the imaginary part of the SS effective action {\it per unit of particle producing surface area}. Kim, Lee and Yoon~\cite{Kim:2009pg} in 2009 presented the real part of the EHS-like action and in doing this extended Nikishov's result. Chervyakov and Kleinert~\cite{Chervyakov:2009bq,Chervyakov:2018nr} re-derived Nikishov's result in a manner that allows for temperature representation~\cite{Muller:1977mm} of the effective action. Gies and Klingmuller~\cite{Gies:2005bz} studied the SS vacuum decay rate {\it per unit volume}, relating in this way to the EHS-LCF limit, see also~\cite{Chervyakov:2018nr}. 

We execute in an explicit manner the implicit suggestions in the work of Chervyakov and Kleinert~\cite{Chervyakov:2009bq} and Gavrilov and Gitman~\cite{Gavrilov:2015yha}: By using the SS there must be a smooth transition between the EHS/LCF-limit to the KP sharp edge potential limit. We introduce prior work on the single SS spontaneous production in \rs{SauterBackground}; in \rs{KPlimSection} we evaluate the KP limit of the imaginary component of effective action explicitly. The KP limit is obtained from a SS with a fixed height and vanishing step width, allowing computation of pair production in the corresponding $\delta$-function-like spiked electric fields. To connect to the opposite limit of smoothly varying fields, we obtain the domain of validity of the EHS-LCF method by comparison with the exact SS results for spin-$1/2$ and spin-0: We compare their imaginary components of effective action in \rs{SectCompare}, and their respective spectra of produced particles in \rs{sectionSpectrum}. We compare the spin-$1/2$ with the spin-0 particle production rates in \rs{compare012}. 

Motivated by a long-standing insight about the difference in the behavior of bound state solutions between spin-$1/2$ and spin-$0$ cases~\cite{Schiff:1940zz}, we explore the case of two opposite sign Sauter steps creating a potential well in \rs{2Sauter}. For sharp-walled wells and spin-$0$ particles, instead of crossing the mass gap, bound states representing both particles and antiparticles meet within the energy gap separating these solutions. Spontaneous particle production is thus affected for spin-$0$, being in this case sensitive to the sharpness of the Sauter steps. 

A physics motivated regularization of the singular behavior of the sharp-walled potential wells is the back-reaction of the condensates of spin-$0$ particle-antiparticle pairs forming a neutral condensate~\cite{Klein:1974gp}.  There is no immediate particle emission, contrary to the spin-$1/2$ case, as pairs are retained in the condensate. On the other hand, when at least one side of the well is sufficiently smooth a charged condensate forms within the well~\cite{Klein:1976xj}, allowing spontaneous particle production phenomena akin to the case of spin-$1/2$ where it can proceed for smooth and sharp walled potentials. 

In concluding remarks \rs{openquest} we present a brief summary of our theoretical achievements, discuss necessary extension of our work before it can be realistically used in possible experimental applications, and address open questions.

 \section{Prior work} 
 \label{SauterBackground}

 Our work builds on many contributions spanning nearly a century, and a coherent presentation requires a brief overview of methods, results and their physical significance:
 
\subsection{Single Sauter step} 
We consider the potential step 
\begin{align}\label{SSpot}
A^\mu=&\;(V_1\;,\; \vec 0)
\;,\qquad
V_1=\frac{\mE_0 L}2\tanh[2z/L]
\;,
\end{align}
where subscript {\lq}1\rq\ denotes a single step, and step height
\begin{align}
\label{Amuform}
\Delta V=&\;V_1(+\infty)-V_1(-\infty)=\mE_0L
\;.
\end{align}
The electric field has one $z$-direction component
\begin{align}
\label{Eform}
\mE=-\partial_z V_1=
-\mE_0 \mathrm{sech}^2[2z/L]\underset{L\to 0}{\longrightarrow} -\frac{\Delta V}{2}\delta(z)
\;,
\end{align}
where $\mE_0=\mE(0)$ is the field at the center of the step. The limit $L\to 0$ is as indicated a $\delta$-function. The SS requires the presence of a dipole charge density
\begin{align}
\rho=&\;
\partial_z \mE=
\frac{4\mE_0}L\frac{\tanh[2z/L]}{\cosh^2[2z/L]}
=-\frac{8 V_1}{L^2}\,\frac{\mE}{\mE_0}
\;,
\end{align}
which behaves like a $\delta^\prime$-function. This behavior agrees with that of a self-consistent potential step obtained with back-reaction effects in charged vacuum solutions~\cite{Muller:1974fh,Madsen:2008vq}.

The Sauter step potential has an exact solution to the Dirac equation~\cite{Sauter:1932gsa}, allowing {\lq}Klein paradox\rq\ pair production under the condition
\begin{align}
\label{EthrEH}
e\Delta V=e\mE_0 L > 2m
\;.
\end{align}
We obtain the vacuum decay rate by pair production evaluating the vacuum persistence probability $P$
\begin{align}
P=|e^{i\Gamma}|^2
\;,
\end{align}
which is less than unity only when when $\mathrm{Im}[\Gamma]>0$. We recall that the vacuum decay rate can be related to the rate of pair production, see~\cite{Cohen:2008wz,Labun:2008re}. 
 
\subsection{Imaginary part of effective action in a single Sauter step} 
\label{ImSSbackground}

The physically relevant quantity in the study of a sharp localized potential step is the action density $\mL$ per unit of time $T$ and \emph{surface} $a_\perp$, instead of the usual action per time and volume: $\mL={\Gamma}/{a_\perp T}$. 
The vacuum decay rate probability $w$ per unit of $a_\perp T$ follows from
\begin{align}
2\mathrm{Im}[\Gamma]\equiv a_\perp T w\;,\quad
w=2\mathrm{Im}[\mL] 
\;.
\end{align}
We use here the results for the imaginary parts of effective action $\mathrm{Im}[\mL]$ presented by Kim~\cite{Kim:2009pg}. 
The two cases of interest are:
\begin{itemize}
\item 
Spin $1/2$: The imaginary part of effective action $\mL^{1/2}_{\mathrm{SS}}(\Delta V,m,L)$, from Eq.\,(22) of~\cite{Kim:2009pg}
\begin{align}
\label{KimImexact}
\mathrm{Im}[\mL^{1/2}_{\mathrm{SS}}]=
-\int_D
\frac{d\omega d^2k_\perp}{(2\pi)^3}
\ln
\Big[\frac{\sinh[\pi\Omega_+/2]\sinh[\pi\Omega_-/2]}
{\sinh[\pi\Delta_+/2]\sinh[\pi\Delta_-/2]}\Big]
,
\end{align}
\vskip -0.5cm
and
\item
Spin-$0$: The imaginary part of effective action $\mL^{0}_{\mathrm{SS}}(\Delta V,m,L)$, from Eq.\,(23) of~\cite{Kim:2009pg}
\begin{align}
\label{KimImexact0}
2\mathrm{Im}[\mL^0_{\mathrm{SS}}]=\!\!
\int_D\frac{d\omega d^2k_\perp}{(2\pi)^3}\ln
\Big[\frac{\cosh[\pi\Omega_+/2]\cosh[\pi\Omega_-/2]}
{\cosh[\pi\Delta_+/2]\cosh[\pi\Delta_-/2]}\Big]
\;,
\end{align}
\vskip -0.5cm
where:
\vskip -0.5cm
\end{itemize}
\begin{align}
\label{expterms}
\Omega_\pm(\Delta V,m,L)=&\;
L(k_{z+}+k_{z-})/2 \pm 2\lambda_\sigma\;,
\nonumber \\ 
\Delta_\pm(\Delta V,m,L)=&\;
L(k_{z+}-k_{z-})/2 \pm 2\lambda_\sigma\;,
\nonumber \\ 
k_{z\pm}(\Delta V,m,L)=&\;\sqrt{(\omega\mp e\Delta V/2)^2-m^2-k_\perp^2} \;,
\nonumber \\ 
\lambda_\sigma=&\;
\sqrt{(e\Delta VL/4)^2-(1-2|\sigma|)^2/4}
\;,
\end{align}
and spin-$1/2$ spin projections $\sigma=\pm1/2$, and for spin $0$ $\sigma=0$. 
As shown in \rf{Fig1}, the domain of integration $D$ in \req{KimImexact} describes the region in which particle states of energy $\omega$, with transverse momentum $k_\perp$, are capable of tunneling through the ($2m$) mass gap 
\begin{align}
\label{intlims}
\int_D d\omega d^2k_\perp =\int_{-(e\mE_0L/2-m)}^{e\mE_0L/2-m} \!\!\!\!\! d\omega
\int_0^{\sqrt{(|\omega|-e\mE_0 L/2)^2-m^2}} \!\! \!\!\! d^2k_\perp 
\;.
\end{align}

%
\begin{figure}[ht]
\center
\includegraphics[width=0.99\columnwidth]{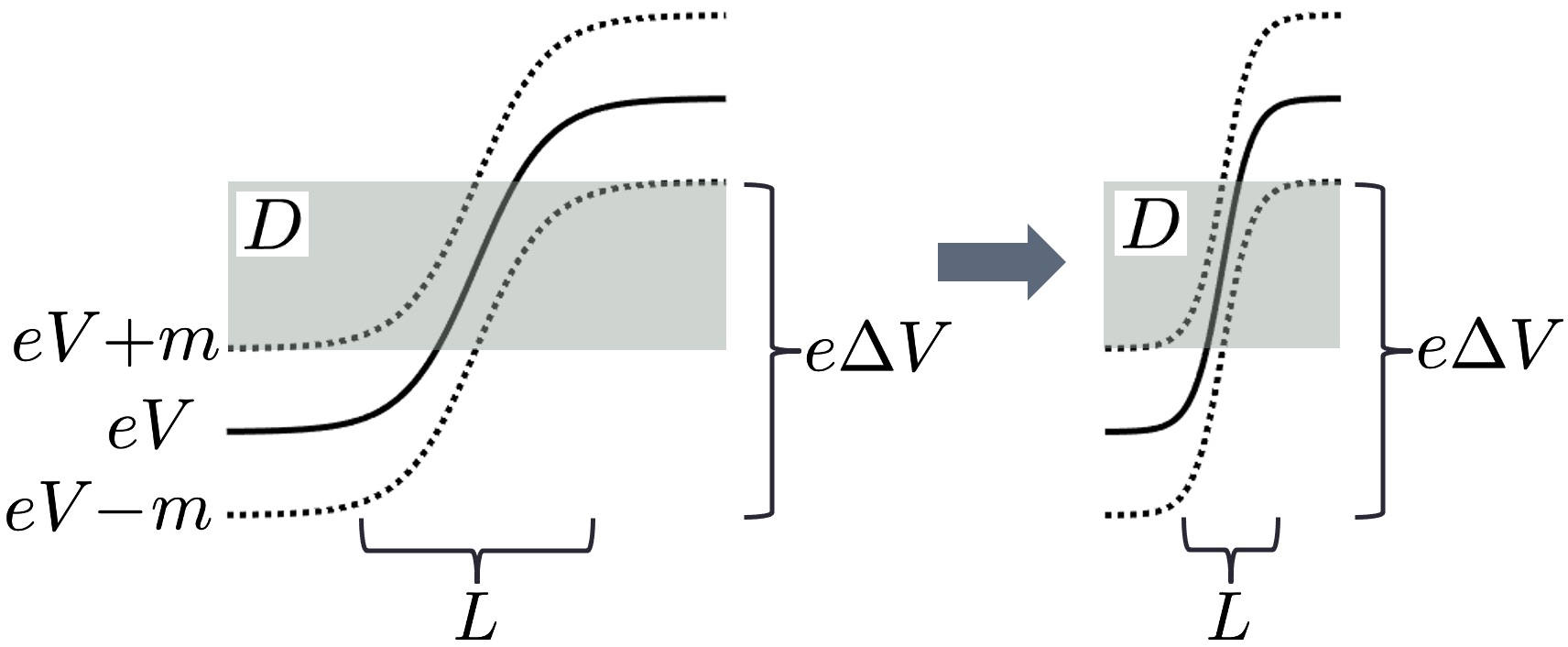}
\caption{\label{Fig1}Klein paradox limit: $L\to0$ as potential step height $e\Delta V=e\mE_0L$ stays fixed. The domain of tunneling states $D$ is unchanged.} 
\end{figure}
%
%

%
\begin{centering}
\begin{figure*}
\includegraphics[width=0.9\textwidth]{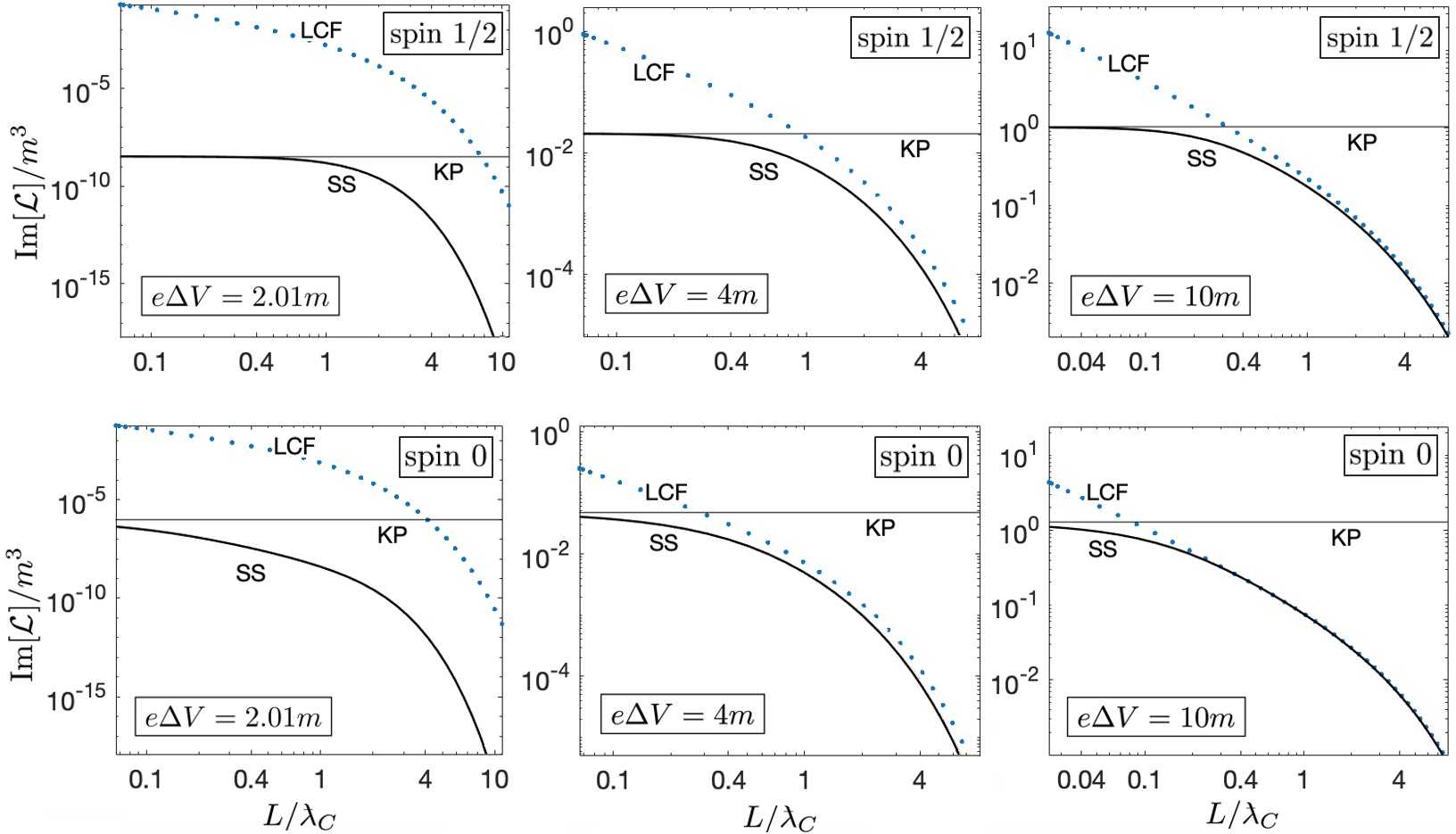}
\caption{\label{Fig3a} Solid lines are the complete SS imaginary action for the $2.01m$, $4m$, and $10m$ potential step heights (from left to right). 
Spin-$1/2$ results are shown in the top row, \req{KimImexact}, and spin-$0$ in the bottom row, \req{KimImexact0}. EHS-LCF approximation are the dotted lines, \req{EHLCF} and \req{EHLCF0} respectively. The thin horizontal line indicates the maximum rate obtained in the KP limit, \req{KleinIm} and \req{KPscalar}, respectively. } 
\end{figure*}
\end{centering}
%
%

The Klein paradox limit is obtained by taking the limit $L\to 0$ in \req{KimImexact}, see \rf{Fig1}. In the opposite limit we obtain the known form of EHS effective action following Nikishov~\cite{Nikishov:1970br}, see also~\cite{Kim:2009pg} and~\cite{Chervyakov:2009bq}, evaluating \req{KimImexact} in the $L\to\infty$ limit: As $L$ increases, the potential step height increases, while the electric field $\mE_0$ stays quasi-constant in an ever larger space domain $V_\mathrm{EHS}\propto L\times a_\perp $; thus we must also find $\mathrm{Im}[\mL^{1/2}_{\mathrm{SS}}]\propto L$.

 \section{Results for pair production in a single Sauter step} 
 \label{newSection}
\subsection{Klein Paradox limit} 
\label{KPlimSection}
\subsubsection*{Spin-0}
The spin-0 case has been treated before~\cite{Chervyakov:2009bq}. We review this case with the objective to prepare the evaluation of the KP limit for spin-$1/2$. 

The KP limit of spin-$0$ SS action is obtained from a fixed potential step height $\Delta V$ with vanishing $L$:
\begin{align}
\lim_{L\to0}
\mathrm{Im}[\mL^{0}_{\mathrm{SS}}(\Delta V,m,L)]
=\mathrm{Im}[\mL^{0}_{\mathrm{KP}}(\Delta V,m)]
\;.
\end{align}
We follow the steps of~\cite{Kim:2009pg,Chervyakov:2009bq} and solve the Bogoliubov coefficients for a step function potential
\begin{align}
\label{Vstep}
V_1=\frac{\mE_0 L}2(\Theta(z)-1/2)
\;.
\end{align}
The momentum states $k_{z+},k_{z-}$ in \req{expterms} apply to $z>0$ and $z<0$, respectively. The Klein-Gordon solutions to the step function potential \req{Vstep}
\begin{align}
\phi_{z<0}=&\;Ae^{ik_{z-}z}
\!+Be^{-ik_{z-}z}
\!,\quad
\phi_{z>0}=
e^{ik_{z+}z}
\;.
\end{align}
From the continuity of $\phi$ and its $\phi^\prime$ at $z=0$, the Bogoliubov coefficient 
\begin{align}
\frac AB=\frac{k_{z-}+k_{z+}}{k_{z-}-k_{z+}}
\;,
\end{align}
and summing the states as in Eq.\,(8) of~\cite{Kim:2009pg},
\begin{align}
\label{KPscalar}
\mathrm{Im}[\mL^0_{\mathrm{KP}}(\Delta V,m)]=&\;
\int_D
\frac{d\omega d^2k_\perp}{(2\pi)^3}\mathrm{Re}[\ln[A/B]]
\nonumber \\
=&\;
\frac12\int_D\frac{d\omega d^2k_\perp}{(2\pi)^3}
 \ln \Big[\frac{(k_{z-}+k_{z+})^2}{(k_{z-}-k_{z+})^2}\Big]
\;.
\end{align}

\subsubsection*{Spin-$1/2$}

For spin-$1/2$ particles we obtain the KP limit:
\begin{align}
\lim_{L\to0}
\mathrm{Im}[\mL^{1/2}_{\mathrm{SS}}(\Delta V,m,L)]
=\mathrm{Im}[\mL^{1/2}_{\mathrm{KP}}(\Delta V,m)]
\;.
\end{align}
We solve the Bogoliubov coefficients for the step function potential \req{Vstep} using the Weyl Dirac solutions: 
\begin{align}
\psi_{z<0}=&\; A\begin{pmatrix}
\phantom{\frac{\omega+e\Delta V/2-2k_{z-}\sigma}m}\ \chi^\sigma \\ 
\frac{\omega+e\Delta V/2-2k_{z-}\sigma}m\ \chi^\sigma\end{pmatrix}
e^{ik_{z-}z}
\nonumber \\
+&\;
B\begin{pmatrix}
\phantom{\frac{\omega+e\Delta V/2+2k_{z-}\sigma }m}\ \chi^\sigma \\ 
\frac{\omega+e\Delta V/2+2k_{z-}\sigma }m\ \chi^\sigma\end{pmatrix}
e^{-ik_{z-}z}
\;,
\nonumber \\[5pt]
\psi_{z>0}=&\;
\begin{pmatrix}
\phantom{\frac{\omega-e\Delta V/2-2k_{z+}\sigma }m}\ \chi^\sigma \\ 
\frac{\omega-e\Delta V/2-2k_{z+}\sigma }m\ \chi^\sigma\end{pmatrix}
e^{ik_{z+}z}
\;,
\end{align}
with 2-spinor $ \chi^\sigma$. Continuity at $z=0$ gives
\begin{align}
\frac AB=\frac{2\sigma(k_{z-}+k_{z+})+e\Delta V}{2\sigma(k_{z-}-k_{z+})+e\Delta V}
\;.
\end{align}
 Plugging again into Eq.\,(8) of~\cite{Kim:2009pg},
\begin{align}
\label{KleinIm}
\mathrm{Im}[\mL^{1/2}_{\mathrm{KP}}(\Delta V,m)]=&\;
(-1)^{2|\sigma|}\!\!\!
\int_D\!
\frac{d\omega d^2k_\perp}{(2\pi)^3}
\!\!\!\sum_{\sigma}^{\pm 1/2}\!\! \mathrm{Re}[\ln[A/B]]
\nonumber \\
=&\;
-\int_D\frac{d\omega d^2k_\perp}{(2\pi)^3}\ln
\Big[\frac{\tilde\Omega_+\tilde\Omega_-}{\tilde\Delta_+ \tilde\Delta_-}\Big]
\;,
\end{align}
where the following quantities depend only on step height and mass:
\begin{align}
\label{Lindep}
\!\!\!\tilde\Omega_\pm(\Delta V,m)=&\;
\Omega_\pm/L=
(k_{z+}+k_{z-}\pm e\Delta V)/2\;,
\nonumber \\
\!\!\!\tilde\Delta_\pm(\Delta V,m)=&\;
\Delta_\pm/L=
(k_{z+}-k_{z-}\pm e\Delta V)/2
\;.
\end{align}

We present KP imaginary actions for both spin-0 and spin-$1/2$ in \rf{Fig3a}, alongside the exact SS results, for the potential step heights $e\Delta V=2.01m$, $4m$ and $10m$. The KP limit \req{KleinIm} (thin horizontal line) is independent of $L$, and provides a finite upper bound to the imaginary action. The exact SS result \req{KimImexact} (solid line, $L$-dependent) is always below the KP limit, while the EHS-LCF approximation (dotted line) exceeds it. This upper bound to production in static fields is reminiscent of the saturation of pairs produced at sufficiently large oscillation frequencies of time-dependent fields~\cite{Krajewska:2018lwe}.

\subsection{Validating the Euler-Heisenberg-Schwinger locally constant field approximation} 
\label{SectCompare}

%
\begin{figure}
\includegraphics[width=0.99\columnwidth]{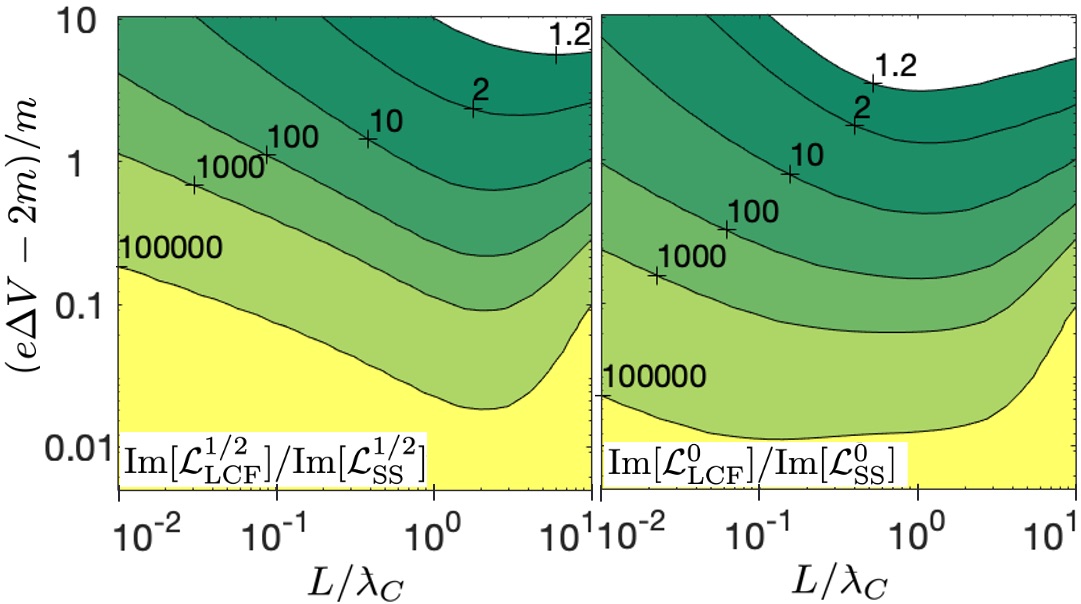}
\caption{ \label{FigCompare} For spin-$1/2$ (left) and spin-0 (right), ratios of the imaginary EHS-LCF action, \req{EHLCF} and \req{EHLCF0}, to the Sauter result, \req{KimImexact} and \req{KimImexact0}, respectively.} 
\end{figure} 
%

The EHS-LCF approximation arises from the EHS analytical result upon integration over $z$: For spin $1/2$
\begin{align}
\label{EHLCF}
\mathrm{Im}[\mL^{1/2}_{\mathrm{LCF}}]=&\; 
\int dz\frac{e^2\mE^2(z)}{8\pi^3}\sum_{n=1}^\infty\frac{e^{-n\pi m^2/e\mE(z)}}{n^2}
\;,
\end{align}
and for spin $0$
\begin{align}
\label{EHLCF0}
\mathrm{Im}[\mL^0_{\mathrm{LCF}}]=&\; 
-\int dz\frac{e^2\mE^2(z)}{16\pi^3}\sum_{n=1}^\infty(-1)^n\frac{e^{-n\pi m^2/e\mE(z)}}{n^2}
\;.
\end{align}
Only where the electric field $\mE(z)$ of the SS given by \req{Eform} is nonvanishing does a contribution, dominated by the domain of maximum field, arise. The integral diverges in the $L\to0$ KP limit due the quadratic in the field pre-factor.

In \rf{FigCompare} we present the ratios of EHS-LCF to exact SS imaginary actions for spin-$1/2$ and spin-0. We see that in the parameter domain first accessible to experiment, near threshold $e\Delta V\sim2m$, the EHS-LCF approximation is not an acceptable approach to guide the design of future experiments. The domain in which the EHS-LCF approximation is valid is obtained by moving diagonally up the plots; both $L$ and $e\Delta V$ values increasing. The upper right hand corners of the plots (domains within contour = $1.2$) correspond to less than $20\%$ error. We find to our surprise that there is a finite optimal $L$, dependent on step height $e\Delta V$, at which the EHS-LCF is closest to the exact SS imaginary action.

\subsection{Spectrum of particles produced}
\label{sectionSpectrum}

%
\begin{figure}[ht]
\includegraphics[width=0.99\columnwidth]{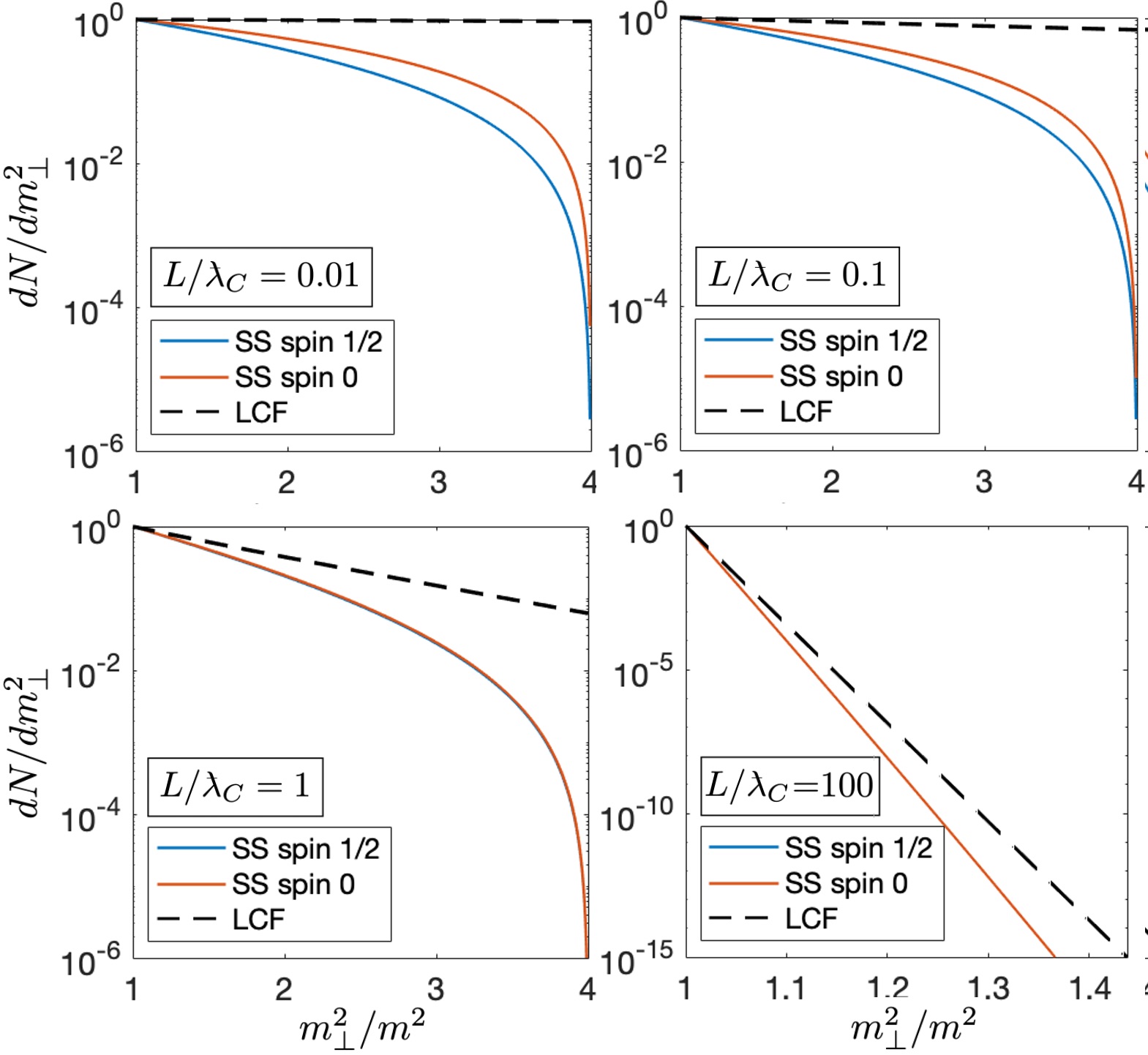}
\caption{ \label{FigSpecrtum} Comparison of the SS $m_\perp^2$ spectra \req{spectrumexact} (solid lines for spin 0 (red) and spin $1/2$ (blue)) with the EHS-LCF approximate spectra, \req{spectrumEHS}, both 
normalized (to 1). Four different step widths ($L/\lambdabar_C=10^{-2},10^{-1 }, 1, 100$), all with step height $e\Delta V=4m$ are shown. For $L/\lambdabar_C>1$, the spin-0 and spin-$1/2$ spectra overlap.}
\end{figure}
%
%

We compute the SS transverse momentum spectra using the mean number of pairs produced~\cite{Nikishov:1970br,Kim:2009pg}
\begin{align}
N_{\omega k_\perp \sigma}= \frac{2\sinh[\pi L k_{z+}/2]\sinh[\pi L k_{z-}/2]}
{\cosh[2\pi\lambda_\sigma]+(-1)^{2\sigma}\cosh[\pi L (k_{z+}-k_{z-})/2]}
\;.
\end{align}
Summing over momentum and spin states, the pair production rate per unit cross sectional area and time
\begin{align}
\label{Nsautertotal}
N_{\mathrm{SS}}=&\;
\sum_\sigma\int_D
\frac{d\omega d^2k_\perp}{(2\pi)^3}
N_{\omega k_\perp \sigma}
\;,
\end{align}
where the domain $D$ of tunneling states is rewritten as
\begin{align}
\label{mperpDomain}
&\;\int_D d\omega d^2k_\perp =
2\pi\int_{m^2}^{m_{\perp\mathrm{max}}^2} \!\!\! dm^2_\perp 
\int_{0}^{e\mE_0L/2-m_\perp}\!\!\! d\omega\;,
\\[7pt] 
\nonumber
&\;m_\perp^2=m^2 +k_\perp^2\;,\quad
m_{\perp\mathrm{max}}^2=(e\mE_0 L/2)^2
\;,
\end{align}
equivalent to \req{intlims} with exchanged order of integration, and with $k_\perp$ written in terms of $m_{\perp}$. With the help of \req{mperpDomain} we can write for \req{Nsautertotal}
 \begin{align}
N_{\mathrm{SS}}=&\;
\int_{m^2}^{m_{\perp\mathrm{max}}^2} \!\!\! dm^2_\perp 
\frac{dN_{\mathrm{SS}}}{dm_\perp^2}
\;,
\end{align}
with the transverse spectrum 
\begin{align}
\label{spectrumexact}
\frac{dN_{\mathrm{SS}}}{dm_\perp^2}=&\;
\frac{1}{(2\pi)^2} \sum_\sigma
\int_{0}^{e\mE_0L/2-m_\perp}\!\!\! d\omega
N_{\omega k_\perp \sigma}
\;.
\end{align}
The exact transverse spectrum \req{spectrumexact} has a maximum value of $m_\perp$ beyond which no particle is produced. Such a finite upper $m_\perp$ limit does not arise in the EHS-LCF approximation~\cite{Cohen:2008wz,Labun:2008re} 
\begin{align}
\label{spectrumEHS}
\frac{dN_{\mathrm{LCF}}}{dm_\perp^2}=
\sum_\sigma\int_{-\infty}^\infty dx \frac{e\mE(x)}{8\pi^2} e^{-\pi m_\perp^2 /e\mE(x)}
\;.
\end{align}
In the EHS limit of SS with $L\to\infty$ at a fixed field strength $\mE_0=\Delta V/L$, see end of \rs{ImSSbackground} for details, \req{spectrumexact} and \req{spectrumEHS} are asymptotically equal.

In \rf{FigSpecrtum} we plot the transverse momentum spectra as functions of $m_{\perp}^2$ (straight lines correspond to Gaussian spectra). The small momentum ($m_\perp\sim m$) agreement between the exact SS and the EHS-LCF spectra is a consequence of normalization. In comparing the two spectra we are thus evaluating their distributions, comparing the relative suppression of large momentum states. Due to the need to satisfy the upper limit $m_{\perp\mathrm{max}}^2=4m^2$, \req{mperpDomain} we find that the exact SS spectra with $L/\lambdabar_C=0.1,0.01$ drop much faster than the Gaussian EHS-LCF distributions. At larger $L$, see $L/\lambdabar_C=100$ example, the SS spectrum becomes Gaussian, yet the distribution remains more narrow than (below) the EHS result.

\subsection{Comparison between spin-$1/2$ and spin-0}
\label{compare012}
%
\begin{figure}[ht]
\center
\includegraphics[width=0.8\columnwidth]{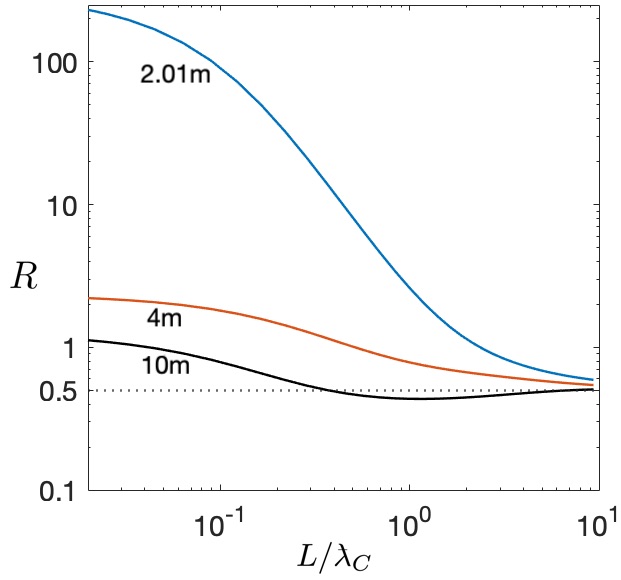}
\caption{\label{constmassplot} Ratio $R$ of the spin-$0$ and spin-$1/2$ imaginary actions \req{R0to12}, for $e\Delta V=2.01m$, $4m$ and $10m$.} 
\end{figure}
%
%

We now turn to comparison of spin-0 with spin-$1/2$ pair production. We define ratio $R$ as
\begin{align}
\label{R0to12}
R=\frac{\mathrm{Im}[\mL^0_{\mathrm{SS}}]}{\mathrm{Im}[\mL^{1/2}_{\mathrm{SS}}]}
\;,
\end{align}
the ratio of exact SS imaginary actions \req{KimImexact0} and \req{KimImexact}. It was noted in work by Gies and Torgrimsson~\cite{Gies:2016coz} that the ratio may be nontrivial. We indeed find, surprisingly to us, that the spin-$0$ production dominates the spin-$1/2$ rate in the KP limit in the proximity of the pair threshold ($e\Delta V=2.01m$), despite spin multiplicity, \rf{constmassplot}. In the large $L$ EHS-LCF limit (dotted line), the spin-$0$ contribution becomes half the spin-$1/2$ rate.

\section{Well potential made of two Sauter steps} 
\label{2Sauter}

We consider field configurations satisfying condition
\begin{align}
\label{localcondition}
\int_{-\infty}^\infty dz \mE(z)=0 
\;.
\end{align}
\req{localcondition} guarantees that the field does not create an asymmetry between particle and antiparticle spectra; see~\cite{Krajewska:2018lwe} for analogous time-dependent condition. A potential satisfying \req{localcondition} consists of at least two potential steps forming a potential well asymptotically normalized to same value (typically 0). 

To evaluate pair production in such fields we begin recalling the known problem with solutions of the spin-$0$ relativistic wave equation: Not all potential well configurations allow for spin-$0$ bound states to \lq dive\rq, with state binding crossing the $2m$ energy gap~\cite{Rafelski:1976ts,Schiff:1940zz,Klein:1974gp,Klein:1976xj}. For elementary book discussion, see p. 59-65 and p. 499-503 of \cite{Greiner:1985ce}. 

For the spin-$1/2 $ case with gyromagnetic ratio $g=2$, the states always dive through the mass gap, irrespective of the steepness of the two sides of the well, p. 59-65 of \cite{Greiner:1985ce}. However, when $g\ne 2$, see Ref.\,\cite{Steinmetz:2018ryf}, and as noted for the spin-$0$ case, which for spin-$1/2 $ corresponds to $g=0$, the structure of bound states is sensitive to the shape of the well. We focus here on the case of spin-$0$ particles.

\begin{figure}[ht]
\center
\includegraphics[width=0.9\columnwidth]{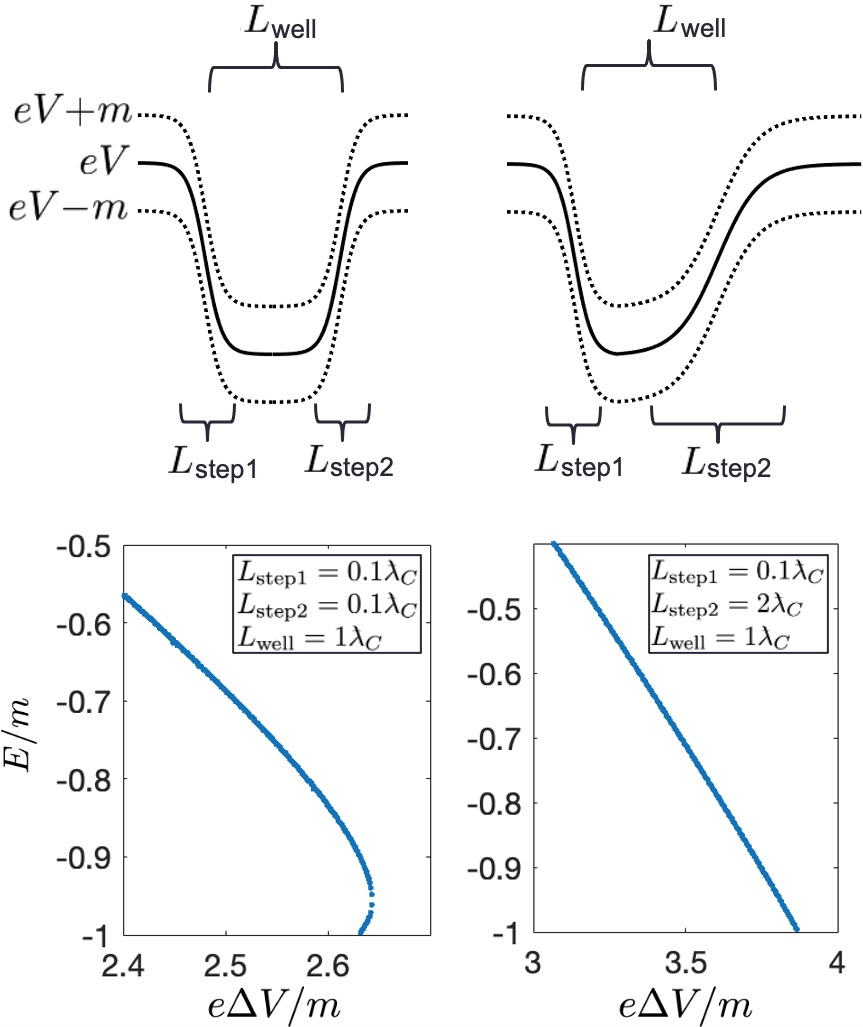}
\caption{\label{asymm} Top: two Sauter steps forming symmetric (LHS) and antisymmetric (RHS) wells. Bottom: Corresponding eigenvalues for the lowest spin-$0$ bound state ($p_\perp=0$) as a function of potential well depth with fixed width of $1\lambdabar_C$: Only the RHS allows direct particle emission.} 
\end{figure}

We consider two static 1D $\tanh$-like SS potentials of opposite sign as shown in Fig.\,~\ref{asymm}:
\begin{align}
\label{Vwell}
V_2=\frac{\Delta V}2\!\!\left(\!\tanh\left[\frac{2z-L_{\mathrm{well}}}{L_{\mathrm{step}1}}\right]
-\tanh\left[\frac{2z+L_{\mathrm{well}}}{L_{\mathrm{step}2}}\right]\right)
\;,
\end{align}
where subscript {\lq}2\rq\ denotes two steps (sides) of the well; the steps are displaced by $\pm L_{\mathrm{well}}/2$; and a depth $e\Delta V>2m$ is required for pair production.

%
%
%

We numerically solve the Klein-Gordon equation in the potential well \req{Vwell}, and then plot the bound states as a function of well depth, at fixed widths: On the LHS of Fig.\,~\ref{asymm}, the symmetric well with steep steps $L_{\mathrm{step}1}=L_{\mathrm{step}2}=0.1\lambdabar_C$ exhibits no pair production. The bound state is unable to dive as it joins the negative energy branch~\cite{Schiff:1940zz}. This situation is addressed in Ref.\,\cite{Greiner:1985ce}, see pp. 499-503. The eigenstate emerging from the lower continuum has a negative normalization signaling opposite charge interpretation in the $2^\mathrm{nd}$-quantized theory; discrete eigenstates are normalized according to $\pm 1=2\int dz \phi_n^\dagger (E_n-V)\phi_n$, for details of 2nd quantization and related (neutral) Bose condensation see Ref.\,\cite{Klein:1974gp}.
 
When the positive and negative energy branches join and disappear from the spectrum, the eigenmodes no longer create a complete set of states, meaning the Klein-Gordon (quantum) Hamilton operator has lost self-adjointness.  However, operators in physics representing physical observables and in particular, the energy, must not only be Hermitian but must also be self-adjoint in order for the time evolution of the system to be unitary, {\it i.e\/} to preserve probability. This point was made in the present context over 70 years ago in Case's work~\cite{Case:1950an}, with the full authority of his mentors Pauli, Oppenheimer, Von Neumann.
The usually followed solution when physical systems in presence of strong fields develop this type of singularity, also noted by Case ({\it loc.cit.\/}), is to establish in first step a physics motivated regularization of the (Hamiltonian) operators which fail self-adjointness test.

It goes without further discussion that in the spin-0 square well case, the complex eigenvalues found mathematically for operators that are not self-adjoint have no physical meaning and certainly are not applicable to context of pair production in the same way as the imaginary part of effective action is. In the latter case we choose in an approximation scheme to work in a sub-space of the full Hilbert space ({\it i.e.\/} at first omitting states with particle pairs). Time evolution in this physical subspace is meaningful even if a transition rate (imaginary part of action) arises, connecting to states containing pairs. This situation is fundamentally different from a study of non-self adjoint operators as these do not span a Hilbert space, and thus there is no physically meaningful subspace.

In our present context we can \lq{}regularize\rq\ the spin-0 square well to restore self-adjointness by making one side smooth: The RHS of Fig.\,~\ref{asymm} shows an asymmetric well where $L_{\mathrm{step}1}=0.1\lambdabar_C$ and $L_{\mathrm{step}2}=2\lambdabar_C$. The bound state crosses the mass gap, and we can take large $L_{\mathrm{well}}$ and $L_{\mathrm{step}2}$ to recover the single SS results for particle production, with spin-$0$ production exceeding the spin-$1/2$ rate as demonstrated in \rf{constmassplot}.

This result implies that for a well potential, no imaginary part of spin-$0$ effective action exists when both sides are taken to the KP sharp edge limit. Clearly this situation deserves more in-depth study of the real part of effective action, which changes in discontinuous fashion each time particle and antiparticle states meet in the mass gap. We note that in the large $L_{\mathrm{well}}$ limit, the joining point (between $\pm$ energy branches) asymptotically approaches the threshold for pair production ($-m$), but never crosses it. In such limiting cases one wonders about the effect of decoherence over much larger (classical) distances. It seems necessary to revert in such a case to the single step result, since {\lq}on the other side\rq\ of the well we encounter a smooth effective potential step created by progressive decoherence of quantum states, instead of a distant sharp step.

\section{Summary and Conclusions}
\label{openquest}
\subsection{Theoretical results}
We have established a relation between the Euler-Heisenberg-Schwinger and the spontaneous particle production from a potential well, creating in explicit computation an understanding of when the application of the EHS-locally-constant-field rate of pair production is allowable. Moreover, the following insights developed in this work extend prior effort on spontaneous particle production: \\

{\bf 1.}\;\;
\emph{The physical interpretation of the sharp-edge KP limit of spin-$1/2$ effective action requires consideration of the vacuum decay rate per unit-surface area:} 

\noindent 
We have clarified the spin-$1/2$ vacuum spontaneous particle production in strong fields, and obtained a finite \emph{upper bound per unit-surface} to the particle emission rate depending only on the potential step height $e\Delta V>2m$, \rf{Fig3a} of \rs{KPlimSection}. Materialization of virtual particles in the presence of the sharp SS is fastest as expected, since the tunneling width vanishes. The unexpected result is that with vanishing width of the barrier the rate per unit surface is finite. This insight contrasts with the evaluation per unit volume~\cite{Chervyakov:2018nr}. 
 
{\bf 2.}\;\; 
\emph{We quantify the domain of validity of the EHS-LCF approximation:}
 
The domain of optimal step width $L$ and height $e\Delta V$, where the EHS-LCF imaginary action is closest to the exact SS rate, surprisingly, spans a finite range of values in $L$. As an example, in \rf{FigCompare} of \rs{SectCompare} we find that for the domain $2m<e\Delta V<5m$, EHS-LCF is closer to the exact SS result when $L= 2\lambdabar_C$ than when $L= 10\lambdabar_C$. Thus any application of EHS-LCF to static localized electric fields must take into account the extraordinary cases where stronger, sharper applied fields are more accurately represented than weaker, smoother fields spanning larger spatial domains.

{\bf 3.}\;\; \emph{We demonstrate additional (compared to EHS) suppression of the produced particle spectra at high momentum: }

\noindent For time independent applied potentials of any type, sharp or smooth, the exact SS solution amplifies the intrinsic contradiction with spontaneous particle production. All spectra known to be occurring in nature have exponential or power law distributions, and never Gaussian as EHS suggests. The exact SS results are always sub-Gaussian, below the EHS spectra, \rf{FigSpecrtum} in \rs{sectionSpectrum}. This result highlights the fact that a theory relying on a time independent potential step for particle production is missing a very important part which converts the gaussian distribution to the thermal-like exponential distributions, where higher momentum states prevail~\cite{Bialas:1999zg,Florkowski:2003mm}.

{\bf 4.}\;\; \emph{We restore self-adjointness to the spin-0 square well case by making one side smooth:}

\noindent Contrary to a single potential step, a well potential described by \req{localcondition} cannot have spin-0 production for sharp edges, due to loss of self-adjointness. By making one side of the well smooth with respect to the Compton wavelength, the eigenstate set becomes complete again (meaning the Hamiltonian is not only Hermitian but also self-adjoint). All bound states with increasing well depth  ultimately cross from the positive to the negative continuum, and spontaneous particle production can occur, \rs{2Sauter}.

\subsection{Connection to experiment}

Our current results may find application: \\

{\bf 1.}\;\;
\emph{Peripheral ultra-relativistic heavy ion collisions~\cite{Bertulani:2005ru,Tuchin:2013ie}:} The electrical fields of colliding ions are a function of collision energy {\it i.e.\/} the Lorentz-factor $\gamma$. Already for moderately relativistic $\gamma\simeq 20$ the shape of the field at the instant of ions passing each other resembles the SS case considered here. The form of the field undergoes a shift as a function of time from the LCF limit for well separated ions, to the KP limit at the instant ions pass each other. In this condition we expect the dominant particle production contribution suitable for probing the KP vacuum decay limit to be \lq heavy\rq\ particles such as muons.\\

{\bf 2.}\;\; 
\emph{Non-relativistic heavy ion collisions~\cite{Rafelski:2016ixr,Popov:2020xmd}:} Spontaneous and induced production of positrons was explored several decades ago in the near adiabatic limit of slowly moving nuclei. We believe that the theory of these processes could be considerably advanced by developing further the context of particle emission through a surface proposed here. This step is essential to achieve a better understanding of positron production rates and their scaling behavior as a function of impact parameter and nuclear charge.\\

{\bf 3.}\;\;
\emph{The ultra-short laser pulse particle production~\cite{Mourou:2006zz,DiPiazza:2011tq}:} As the high-intensity light frontier advances, pulsed laser field configurations will reach the barely critical (for electrons) potential step regime. Our results demonstrate that actual pair production rates may be orders of magnitude smaller compared to locally constant field estimates, see \rf{FigCompare}: For a recent review of such work see~\cite{Dunne:2008kc,Dunne:2009gi}. However, the incorporation of the dynamics of time dependent fields and light pulse motion may be key to more definitive understanding of this experimental case.\\

{\bf 4.}\;\; 
\emph{Astrophysics~\cite{Ruffini:2009hg,Churazov:2020}:} The electron-positron annihilation line at 511 keV is the most prominent spectral feature in the $\gamma$-ray spectrum of the Milky Way, where the origin of positrons remains an open question. Extreme field strengths are predicted for cataclysmic merger and collapse events in astrophysics. Implementation of the here advanced particle production per unit-surface SS model should allow a better estimation of the positron formation yields in such events and hence the expected positron emission, and annihilation radiation intensity.

\subsection{Open questions}

In this work we have considered the case of fixed in space potential configurations. Thus we did not explore the assistance to particle production provided by the eventual time dependence of field configurations~\cite{Schutzhold:2008pz} -- in early literature this effect was called {\lq}induced\rq\ (by time dependence), see Refs.\,\cite{Rafelski:2016ixr,Greiner:1985ce}. Initial steps to accommodate the effect of time dependence for SS action were already made by Narozhnyi and Nikishov~\cite{Nikishov:1970bN}. During the writing of this paper the authors became aware of the recently submitted work on sharp time-dependent steps~\cite{Breev:2021lpn}, extending the exactly solvable class of field profiles such as the Sauter step to asymmetric (single step) configurations. 

Accommodation of time dependence is without doubt a required further development of the results we present to allow application in actual experiments. For the spin-$0$ case, with a sharp-walled well potential, this is a critical step to obtain actual pair production ensuing from the break-up of the condensate formed in this situation. The remaining question is whether the charge neutral pair condensate releases pairs (sudden potential wall removal), or if the pairs self-annihilate (slow wall removal). In addition, when pair production is abundant it is also essential to discuss the effect of back reaction on the produced spectra of particles~\cite{Muller:1974fh,Kluger:1992gb}. 

Another potential future development relates to the analytical properties of the Sauter effective action which differ from the EHS result. The EHS action for constant fields possesses an essential singularity in the imaginary part for a vanishing electrical field. However, this is not the case for fields existing for finite time periods: Bialynicki-Birula, Rudnicki and Wienczek~\cite{BialynickiBirula:2011eg} have demonstrated that even an adiabatic switch-on suffices to assure the absence of an essential singularity for the pair production rate in time-dependent fields. A similar study for spatially varying SS fields would greatly improve understanding of this phenomenon. How the removal of the essential singularity affects the real part of the effective action is also an open question.

The EHS action (both its real and imaginary parts) was shown to possess an effective temperature representation obeying Bose statistics~\cite{Muller:1977mm}. The imaginary part of the SS effective action has been connected to spin statistics~\cite{Chervyakov:2009bq,Chervyakov:2018nr}. Whether this representation holds in general for fields of finite spacetime extent is of considerable interest. 

There is long-standing interest~\cite{OConnell:1968spc, Dittrich:1977ee, Kruglov:2001dp, Angeles-Martinez:2011wpn, Labun:2012jf, Evans:2018kor} in understanding how the effective action depends on the anomalous magnetic moment $g-2$: Generalizing to arbitrary spin $g$-factors is possible in the EHS action by implementing the Klein-Gordon-Pauli equation (second order Dirac equation with $g\neq 2$~\cite{Steinmetz:2018ryf}). 
We applied the same extension in the KP limit of SS studying the bound state structure in the well; spontaneous pair production requires $|g|>1$. This result predicts interesting $g-2$ behavior for SS effective action, a topic of potential future interest. 

We are also interested in a better understanding of the non-perturbative vacuum response of spin-$1/2$ ($e^+-e^-$) pair fluctuations in strong rapidly moving EM-field configurations: This is in our opinion of essential importance for fully understanding the interaction of intense light pulses with the EM-vacuum. We expect a wealth of non-trivial behavior: We recall that the available real part of SS effective action~\cite{Kim:2009pg} relies upon the imaginary part being nonzero, {\it i.e.} the potential step being capable of producing pairs: $e\Delta V>2m$. We further know that for small $|g|<1$, and in the spin-$0$ ($g=0$) case, for sharp-walled wells, the imaginary part of the action can be null for arbitrary step heights (no particle production), including the supercritical $e\Delta V>2m$. Understanding of the real part of the effective action where its imaginary part vanishes could rely on the analytical structure of results here considered.



\begin{thebibliography}{99}
\bibliographystyle{elsarticle-num}
 
\bibitem{Sauter:1932gsa}
F.~Sauter,
\lq\lq Zum \lq Kleinschen Paradoxon\rq,\rq\rq\
Z. Phys. \textbf{73} (1932), 547-552
doi:10.1007/BF01349862

\bibitem{Popov:2020xmd}
R.~V.~Popov, V.~M.~Shabaev, D.~A.~Telnov, I.~I.~Tupitsyn, I.~A.~Maltsev, Y.~S.~Kozhedub, A.~I.~Bondarev, N.~V.~Kozin, X.~Ma and G.~Plunien, T.~St\"ohlker, D.~A.~Tumakov, and V.~A.~Zaytsev, 
\lq\lq How to access QED at a supercritical Coulomb field,\rq\rq\
Phys. Rev. D \textbf{102} (2020) no.7, 076005
doi:10.1103/PhysRevD.102.076005
[arXiv:2008.05005 [hep-ph]] 

\bibitem{Rafelski:2016ixr}
J.~Rafelski, J.~Kirsch, B.~M\"uller, J.~Reinhardt and W.~Greiner,
\lq\lq Probing QED Vacuum with Heavy Ions,\rq\rq; In: Schramm S., Schäfer M. (eds) \emph{New Horizons in Fundamental Physics,} pp 211-251, FIAS Interdisciplinary Science Series (Springer 2016).
doi:10.1007/978-3-319-44165-8\_17
[arXiv:1604.08690 [nucl-th]]

\bibitem{Bertulani:2005ru}
C.~A.~Bertulani, S.~R.~Klein and J.~Nystrand,
\lq\lq Physics of ultra-peripheral nuclear collisions,,\rq\rq\
Ann. Rev. Nucl. Part. Sci. \textbf{55} (2005), 271-310
doi:10.1146/annurev.nucl.55.090704.151526
[arXiv:nucl-ex/0502005 [nucl-ex]].

\bibitem{Tuchin:2013ie}
K.~Tuchin,
\lq\lq Particle production in strong electromagnetic fields in relativistic heavy-ion collisions,\rq\rq\
Adv. High Energy Phys. \textbf{2013} (2013), 490495
doi:10.1155/2013/490495
[arXiv:1301.0099 [hep-ph]]

\bibitem{Mourou:2006zz}
G.~A.~Mourou, T.~Tajima and S.~V.~Bulanov,
\lq\lq Optics in the relativistic regime,\rq\rq\ 
Rev. Mod. Phys. \textbf{78} (2006), 309-371
doi:10.1103/RevModPhys.78.309


\bibitem{DiPiazza:2011tq}
A.~Di Piazza, C.~M\"uller, K.~Z.~Hatsagortsyan and C.~H.~Keitel,
\lq\lq Extremely high-intensity laser interactions with fundamental quantum systems,\rq\rq\
Rev. Mod. Phys. \textbf{84} (2012), 1177
doi:10.1103/RevModPhys.84.1177
[arXiv:1111.3886 [hep-ph]] 

\bibitem{Ruffini:2009hg}
R.~Ruffini, G.~Vereshchagin and S.~S.~Xue,
\lq\lq Electron-positron pairs in physics and astrophysics: from heavy nuclei to black holes,\rq\rq\
Phys. Rept. \textbf{487} (2010), 1-140
doi:10.1016/j.physrep.2009.10.004
[arXiv:0910.0974 [astro-ph.HE]]

\bibitem{Churazov:2020}
E.~Churazov, L.~Bouchet, P.~Jean, E.~Jourdain, J.~Kn\"odlseder, R.~Krivonos, J.-P.~Roques, S.~Sazonov, T.~Siegert, A.~Strong, R.~Sunyaev,
\lq\lq INTEGRAL results on the electron-positron annihilation radiation and X-ray \& Gamma-ray diffuse emission of the Milky Way,\rq\rq\
New Astronomy Reviews, \textbf{90} (2020), 101548 
doi:10.1016/j.newar.2020.101548 

\bibitem{Sauter:1931zz}
F.~Sauter,
\lq\lq Uber das Verhalten eines Elektrons im homogenen elektrischen Feld nach der relativistischen Theorie Diracs,\rq\rq\
Z. Phys. \textbf{69} (1931), 742-764
doi:10.1007/BF01339461

\bibitem{Heisenberg:1935qt}
 W.~Heisenberg and H.~Euler,
 \lq\lq Folgerungen aus der Diracschen Theorie des Positrons,\rq\rq\
 Z.\ Phys.\ {\bf 98}, 714 (1936).
 doi:10.1007/BF01343663
 [physics/0605038]
 
\bibitem{Weisskopf:1996bu}
 V.~Weisskopf,
 \lq\lq The electrodynamics of the vacuum based on the quantum theory of the electron,\rq\rq\
 Kong.\ Dan.\ Vid.\ Sel.\ Mat.\ Fys.\ Med.\ {\bf 14}, N6, 1 (1936). 

\bibitem{Schwinger:1951nm}
 J.~S.~Schwinger,
 \lq\lq On gauge invariance and vacuum polarization,\rq\rq\
 Phys.\ Rev.\ {\bf 82}, 664 (1951).
 doi:10.1103/PhysRev.82.664 

\bibitem{Dunne:2008kc}
G.~V.~Dunne,
\lq\lq New Strong-Field QED Effects at ELI: Nonperturbative Vacuum Pair Production,\rq\rq\
Eur. Phys. J. D \textbf{55} (2009), 327-340
doi:10.1140/epjd/e2009-00022-0
[arXiv:0812.3163 [hep-th]]

\bibitem{Dunne:2009gi}
G.~V.~Dunne, H.~Gies and R.~Sch\"utzhold,
\lq\lq Catalysis of Schwinger Vacuum Pair Production,\rq\rq\
Phys. Rev. D \textbf{80} (2009), 111301
doi:10.1103/PhysRevD.80.111301
[arXiv:0908.0948 [hep-ph]]

\bibitem{Klein:1929zz}
O.~Klein,
\lq\lq Die Reflexion von Elektronen an einem Potentialsprung nach der relativistischen Dynamik von Dirac,\rq\rq\
Z. Phys. \textbf{53} (1929), 157
doi:10.1007/BF01339716

\bibitem{Dombey:1999id}
N.~Dombey and A.~Calogeracos,
\lq\lq Seventy years of the Klein paradox,\rq\rq\
Phys. Rept. \textbf{315} (1999), 41-58
doi:10.1016/S0370-1573(99)00023-X

\bibitem{Rafelski:1974rh}
J.~Rafelski, B.~M\"uller and W.~Greiner,
\lq\lq The charged vacuum in over-critical fields,\rq\rq\
Nucl. Phys. B \textbf{68} (1974), 585-604
doi:10.1016/0550-3213(74)90333-2 

\bibitem{Rafelski:1976ts}
J.~Rafelski, L.~P.~Fulcher and A.~Klein,
\lq\lq Fermions and Bosons Interacting with Arbitrarily Strong External Fields,\rq\rq\
Phys. Rept. \textbf{38} (1978), 227-361
doi:10.1016/0370-1573(78)90116-3

\bibitem{Greiner:1985ce}
W.~Greiner, B.~M\"uller and J.~Rafelski, \textit{Quantum Electrodynamics of Strong Fields}, (Springer, Heidelberg, 1985) 594p.

\bibitem{Sveshnikov:2019aqg}
K.~A.~Sveshnikov, Y.~S.~Voronina, A.~S.~Davydov and P.~A.~Grashin,
\lq\lq Essentially Nonperturbative Vacuum Polarization Effects in a Two-Dimensional Dirac-Coulomb System with $ Z >Z_{cr}$: Vacuum Charge Density,\rq\rq\
Theor. Math. Phys. \textbf{198} (2019) no.3, 331-362
doi:10.1134/S0040577919030024

 
\bibitem{Nikishov:1970br}
A.~I.~Nikishov,
\lq\lq Barrier scattering in field theory removal of Klein paradox,\rq\rq\
Nucl. Phys. B \textbf{21} (1970), 346-358
doi:10.1016/0550-3213(70)90527-4

\bibitem{Nikishov:1970bN}
 N.~B.~Narozhnyi and A.~I.~Nikishov,
\lq\lq The simplest process in a pair-producing electric field,\rq\rq\
Yad. Fiz. \textbf{11}, 1072 (1970) [Sov. J. Nucl. Phys. \textbf{11}, 596 (1970)].

\bibitem{Nikishov:2003ig}
A.~I.~Nikishov,
\lq\lq Scattering and pair production by a potential barrier,\rq\rq\
Phys. Atom. Nucl. \textbf{67} (2004), 1478-1486
doi:10.1134/1.1788038

\bibitem{Kim:2009pg}
S.~P.~Kim, H.~K.~Lee and Y.~Yoon,
\lq\lq Effective Action of QED in Electric Field Backgrounds II. Spatially Localized Fields,\rq\rq\
Phys. Rev. D \textbf{82} (2010), 025015
doi:10.1103/PhysRevD.82.025015

 \bibitem{Chervyakov:2009bq}
A.~Chervyakov and H.~Kleinert,
\lq\lq Exact Pair Production Rate for a Smooth Potential Step,\rq\rq\
Phys. Rev. D \textbf{80} (2009), 065010
doi:10.1103/PhysRevD.80.065010
 
\bibitem{Chervyakov:2018nr} 
A.~Chervyakov and H.~Kleinert,
\lq\lq On Electron\textendash{}Positron Pair Production by a Spatially Inhomogeneous Electric Field,\rq\rq\
Phys. Part. Nucl. \textbf{49} (2018) no.3, 374-396
doi:10.1134/S1063779618030036

\bibitem{Gavrilov:2015yha}
S.~P.~Gavrilov and D.~M.~Gitman,
\lq\lq Quantization of charged fields in the presence of critical potential steps,\rq\rq\
Phys. Rev. D \textbf{93} (2016) no.4, 045002
doi:10.1103/PhysRevD.93.045002
 

 \bibitem{Muller:1977mm}
 B.~M\"uller, W.~Greiner and J.~Rafelski,
 \lq\lq Interpretation of External Fields as Temperature,\rq\rq\
 Phys.\ Lett.\ A {\bf 63} (1977) 181.
 doi:10.1016/0375-9601(77)90866-0

\bibitem{Gies:2005bz}
H.~Gies and K.~Klingmuller,
\lq\lq Pair production in inhomogeneous fields,\rq\rq\
Phys. Rev. D \textbf{72} (2005), 065001
doi:10.1103/PhysRevD.72.065001

 \bibitem{Schiff:1940zz}
L.~I.~Schiff, H.~Snyder and J.~Weinberg,
\lq\lq On The Existence of Stationary States of the Mesotron Field,\rq\rq\
Phys. Rev. \textbf{57} (1940), 315-318
doi:10.1103/PhysRev.57.315

\bibitem{Klein:1974gp}
A.~Klein and J.~Rafelski,
\lq\lq Bose Condensation in Supercritical External Fields,\rq\rq\
Phys. Rev. D \textbf{11} (1976), 300
doi:10.1103/PhysRevD.11.300

\bibitem{Klein:1976xj}
A.~Klein and J.~Rafelski,
\lq\lq Bose Condensation in Supercritical External Fields. 2. Charged Condensates, \rq\rq\
Z. Phys. A \textbf{284} (1978), 71
doi:10.1007/BF01433878

\bibitem{Muller:1974fh}
B.~M\"uller and J.~Rafelski,
\lq\lq Stabilization of the Charged Vacuum Created by Very Strong Electrical Fields in Nuclear Matter,\rq\rq\
Phys. Rev. Lett. \textbf{34} (1975), 349
doi:10.1103/PhysRevLett.34.349

\bibitem{Madsen:2008vq}
J.~Madsen,
\lq\lq Universal Charge-Radius Relation for Subatomic and Astrophysical Compact Objects,\rq\rq\
Phys. Rev. Lett. \textbf{100} (2008), 151102
doi:10.1103/PhysRevLett.100.151102
[arXiv:0804.2140 [hep-ph]]

\bibitem{Cohen:2008wz}
T.~D.~Cohen and D.~A.~McGady,
\lq\lq The Schwinger mechanism revisited,\rq\rq\
Phys. Rev. D \textbf{78} (2008), 036008
doi:10.1103/PhysRevD.78.036008
[arXiv:0807.1117 [hep-ph]]

\bibitem{Labun:2008re}
L.~Labun and J.~Rafelski,
\lq\lq Vacuum Decay Time in Strong External Fields,\rq\rq\
Phys. Rev. D \textbf{79} (2009), 057901
doi:10.1103/PhysRevD.79.057901
[arXiv:0808.0874 [hep-ph]]

\bibitem{Krajewska:2018lwe}
K.~Krajewska and J.~Z.~Kami\'nski,
\lq\lq Threshold effects in electron-positron pair creation from the vacuum: Stabilization and longitudinal versus transverse momentum sharing,\rq\rq\
Phys. Rev. A \textbf{100} (2019) no.1, 012104
doi:10.1103/PhysRevA.100.012104
[arXiv:1811.07528 [hep-ph]].

\bibitem{Gies:2016coz}
H.~Gies and G.~Torgrimsson,
\lq\lq Critical Schwinger pair production II - universality in the deeply critical regime,\rq\rq\
Phys. Rev. D \textbf{95} (2017) no.1, 016001
doi:10.1103/PhysRevD.95.016001

\bibitem{Steinmetz:2018ryf}
A.~Steinmetz, M.~Formanek and J.~Rafelski,
\lq\lq Magnetic Dipole Moment in Relativistic Quantum Mechanics,\rq\rq\
Eur. Phys. J. A \textbf{55} (2019) no.3, 40
doi:10.1140/epja/i2019-12715-5
[arXiv:1811.06233 [hep-ph]]

\bibitem{Case:1950an}
K.~M.~Case,
\lq\lq Singular potentials,\rq\rq\
Phys. Rev. \textbf{80} (1950), 797-806
doi:10.1103/PhysRev.80.797

%

\bibitem{Bialas:1999zg}
A.~Bialas,
\lq\lq Fluctuations of string tension and transverse mass distribution,\rq\rq\
Phys. Lett. B \textbf{466} (1999), 301-304
doi:10.1016/S0370-2693(99)01159-4
[arXiv:hep-ph/9909417 [hep-ph]].

\bibitem{Florkowski:2003mm}
W.~Florkowski,
\lq\lq Schwinger tunneling and thermal character of hadron spectra,\rq\rq\
Acta Phys. Polon. B \textbf{35} (2004), 799-808
[arXiv:nucl-th/0309049 [nucl-th]].

\bibitem{Schutzhold:2008pz}
R.~Sch\"utzhold, H.~Gies and G.~Dunne,
\lq\lq Dynamically assisted Schwinger mechanism,\rq\rq\
Phys. Rev. Lett. \textbf{101} (2008), 130404
doi:10.1103/PhysRevLett.101.130404
[arXiv:0807.0754 [hep-th]]

 \bibitem{Breev:2021lpn}
A.~I.~Breev, S.~P.~Gavrilov, D.~M.~Gitman and A.~A.~Shishmarev,
\lq\lq Vacuum instability in time-dependent electric fields. New example of exactly solvable case,\rq\rq\
[arXiv:2106.06322 [hep-th]]


\bibitem{Kluger:1992gb}
Y.~Kluger, J.~M.~Eisenberg, B.~Svetitsky, F.~Cooper and E.~Mottola,
\lq\lq Fermion pair production in a strong electric field,\rq\rq\
Phys. Rev. D \textbf{45} (1992), 4659-4671
doi:10.1103/PhysRevD.45.4659

\bibitem{BialynickiBirula:2011eg}
I.~Bialynicki-Birula and L.~Rudnicki,
\lq\lq Removal of the Schwinger nonanaliticity in pair production by adiabatic switching of the electric field,\rq\rq\
[arXiv:1108.2615 [hep-th]]


\bibitem{OConnell:1968spc}
R.~F.~O'Connell,
\lq\lq Effect of the Anomalous Magnetic Moment of the Electron on the Nonlinear Lagrangian of the Electromagnetic Field,\rq\rq\
Phys. Rev. \textbf{176} (1968), 1433-1437
doi:10.1103/PhysRev.176.1433

\bibitem{Dittrich:1977ee}
W.~Dittrich,
\lq\lq One Loop Effective Potential with Anomalous Moment of the electron,\rq\rq\
J. Phys. A \textbf{11} (1978), 1191
doi:10.1088/0305-4470/11/6/019

\bibitem{Kruglov:2001dp}
S.~I.~Kruglov,
\lq\lq Pair production and vacuum polarization of arbitrary spin particles with EDM and AMM,\rq\rq\
Annals Phys. \textbf{293} (2001), 228-239
doi:10.1006/aphy.2001.6186
[arXiv:hep-th/0110061 [hep-th]]

\bibitem{Angeles-Martinez:2011wpn}
R.~Angeles-Martinez and M.~Napsuciale,
\lq\lq Renormalization of the QED of second order spin-$1/2$ fermions,\rq\rq\
Phys. Rev. D \textbf{85} (2012), 076004
doi:10.1103/PhysRevD.85.076004
[arXiv:1112.1134 [hep-ph]]

\bibitem{Labun:2012jf}
L.~Labun and J.~Rafelski,
\lq\lq Acceleration and Vacuum Temperature,\rq\rq\
Phys. Rev. D \textbf{86} (2012), 041701
doi:10.1103/PhysRevD.86.041701
[arXiv:1203.6148 [hep-ph]]

 \bibitem{Evans:2018kor}
S.~Evans and J.~Rafelski,
\lq\lq Vacuum stabilized by anomalous magnetic moment,\rq\rq\
Phys. Rev. D \textbf{98} (2018) no.1, 016006
doi:10.1103/PhysRevD.98.016006
[arXiv:1805.03622 [hep-ph]]

 
\end{thebibliography}
\end{document}